\documentclass{osa-article}

\journal{oe}

\articletype{Research Article}

\begin{document}

\title{Waveguide lattice based architecture for multichannel optical transformations}

\author{N.N. Skryabin,\authormark{1,2,*} I.V. Dyakonov,\authormark{1} M.Yu. Saygin,\authormark{1} and S.P. Kulik\authormark{1}}

\address{\authormark{1}Quantum Technology Centre, Faculty of Physics, Lomonosov Moscow State University, Leninskie gory 1, building 35, 119991, Moscow, Russia\\
\authormark{2}Moscow Institute Physics and Technology, Institutskiy per. 9, Dolgoprudny, 141701, Russia}

\email{\authormark{*}nikolay.skryabin@phystech.edu} 



\begin{abstract}
We consider coupled waveguide lattices as an architecture that implement a wide range of multiport transformations. In this architecture, a particular transfer matrix is obtained through setting the step-wise profiles of the propagation constants seen by the field evolving in the lattice. To investigate the transformation capabilities, the implementation of a set of transfer matrices taken at random and particular cases of discrete Fourier transform, Hadamard and permutation matrices have been described. Because the waveguide lattices schemes are more compact than their traditional lumped-parameter counterparts, our architecture may be beneficial for using in photonic information processing systems of the future.
\end{abstract}

\section{Introduction}

The linear transformations of multiple optical channels are often required by novel information processing systems, e.g., for unscrambling the field modes~\cite{Annoni2017}, the matrix multiplication routine in photonic Ising machines~\cite{Prabhu2020} and neural networks~\cite{Englund2017, Wetzstein2020} and cryptographic tasks~\cite{CryptoTasks}. In recent years, a great deal of interest to linear multichannel photonics has been shown by the quantum community, where it is a necessary ingredient of quantum information processing systems having the prospect to outperform the classical counterparts in solving certain tasks. Among them are quantum simulators that sample discrete photon~\cite{BosonSampling20photons,Bentivegnae1400255} or continuous variable states of light~\cite{Zhongeabe8770}, and more demanding quantum computers, wherein multichannel transformations are utilized to prepare resource quantum states~\cite{Gimeno-Segovia2015,GubarevPRA} and to implement quantum logic operations~\cite{Starek2016,Qiang2018}.

The established way of implementing programmable multichannel schemes exploits the lumped-parameter approach, in which the schemes come in the form of meshes of building blocks, each of which performs only single discrete function over a subset of channels. There exist several decomposition methods using this approach that enable construction of programmable photonic circuits capable of implementing a range of linear transformations covering all~\cite{Reck1994,Clements16} or almost all possible~\cite{Fldzhyan20,Saygin20} transfer matrices. Most often, the meshes' building blocks are the Mach-Zehnder interferometers (MZIs)~\cite{Reck1994,Clements16}. Robustness to errors in programmable schemes is achieved by getting rid of the MZIs and using instead beam-splitters (BSs) with a variable phase shifts~\cite{Fldzhyan20}, as well as using multichannel static blocks placed alternately with phase shifts layers~\cite{Saygin20}. In this type of multichannel schemes  the bends connecting the building blocks occupy the precious real-estate of the photonic chips, thus, limiting the scaling and miniaturization capatibilities.

In the recent years, waveguide lattices have become the object of active research, since they implement multichannel field evolution that can be mapped to some useful transformations, e.g., angular momentum rotation  \cite{Perez2013}, discrete fractional Fourier transform \cite{Weimann2016}, generator of W-states \cite{Grafe2014} and photonic quantum gates \cite{Lahini2018}. Besides, it is a platform for optical simulation experiments at the scale exceeding the one provided by the photonic circuits \cite{Szameit2013, Tangeaat3174}. Comparing the  waveguide lattices with the ones of lumped-parameters circuits, they are more compact and have lower losses, since the lack of bent segments of waveguides. However, so far, the waveguide lattices were lacking programmability, so that each new transformation would require manufacturing the corresponding static scheme. 

In this work, we study the waveguides lattices, in which programmablity is enabled by the profiles of the propagation constants experienced by the fields in the course of their evolution. We study the transformation capabilities of the lattices by analyzing transformation fidelity for a set of transfer matrices taken at random from a uniform distribution and for particular cases: discrete Fourier transform, Hadamard and permutation matrices have been described.

\section{Methods}

\subsection{Field evolution in the waveguide lattices}

\begin{figure}[htbp]
\centering\includegraphics[width=0.7\textwidth]{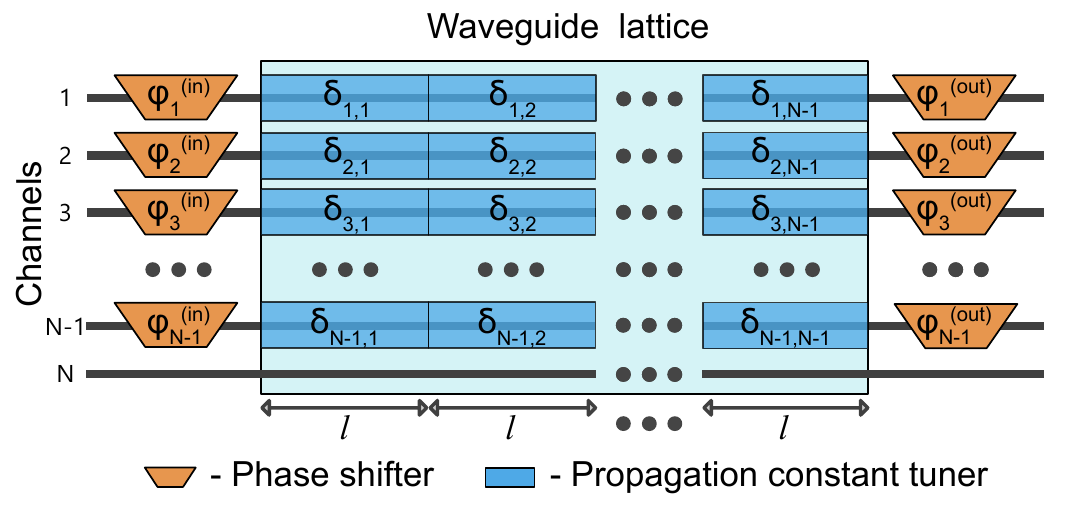}
\caption{The proposed architecture of multichannel linear transformation. The $N$-channel scheme consists of a waveguide lattice and phase shifts placed at the input and output. The lattice is evanescently coupled waveguides with static coupling constants $C_{m,m+1}$ and a variable set of propagation constants defined in a stepwise manner over $N-1$ layers. Here, $\delta_{mj}$ is the propagation constant in waveguide $m$ and layer $j$ relative to the one of the last waveguide, which is assumed constant (see the main text for details), $\varphi^{(in)}_m$ and $\varphi^{(out)}_m$ are variable phase shifts ($m=\overline{1,N-1};j=\overline{1,N-1}$).
}
\label{fig:lattice_scheme}
\end{figure}

The scheme architecture under consideration is depicted in Fig.~\ref{fig:lattice_scheme}. Its main part is the waveguide lattice providing the  multichannel interference required for general linear transformations to take place. The  lattice consists of layers of equal length $l$ that differ by the propagation constants $\beta_m(z)$ experienced by the fields, where $m$ denotes the waveguide and $z$ is the propagation coordinate. We describe a model in which the propagation constants are constant within layers, so that $\beta_m(z)$ is step-wise. Also, phase shifts, $\varphi^{(in)}_m$ and $\varphi^{(out)}_m$, are added at the input and output of the lattice. It is the aim of this work to study the architecture capabilities to implement various multichannel linear transformations by setting  $\beta_m(z)$ and $\varphi^{(in)}_m$ and $\varphi^{(out)}_m$ at fixed value of layer length $l$ for given $N$.

Our analysis of the waveguide lattice to implement multichannel transformations is rested on the following assumptions. Firstly, we assume the nearest-neighbor  interaction justified when the coupling strengths $C_{mn}$ between the waveguides of indices $m$ and $n$ fulfill the condition: $C_{m,m+1}\gg{}C_{m,m+q}$ for $q\ge2$. The second assumption is about the characteristic variance of the refractive indices of the waveguide cores $\delta{}n(z)$, always present due to the necessity to change the propagation constants from layer to layer and from waveguide to waveguide. Namely, we take $\delta{}n(z)$ as having the effect only on the propagation constants $\beta_j(z)$ with negligible effect on the coupling strengths $C_{mn}$, therefore, in the following the coupling strengths $C_{mn}$ will be constant. In particular, this is justified in the case of weakly coupled waveguides, e.g., those fabricated with femtosecond laser writing (see Supplementary of \cite{Heinrich2014}). In addition, we neglect losses usually occuring due to scattering and material absorption, thus, the field evolution is unitary.

Using the coupled mode theory, the following system of equations governs the field amplitudes in the waveguide lattice\cite{Huang94}:
\begin{equation}\label{eqn:coupled_waveguides} 
    i\frac{d\mathbf{A}(z)}{dz} = H(z)\mathbf{A}(z),
\end{equation}
where $\mathbf{A}(z)=(A_1(z)\ldots{}A_N(z))^T$ is the column of field amplitudes at propagation distance $z$, $H(z)$ is the $N\times{}N$ matrix: 
\begin{equation}
    H(z) = \begin{bmatrix}
    \beta_{1}(z) &C_{12} &\ldots &0 \\
    C_{21} & \beta_{2}(z) &\ldots &\vdots \\
    \vdots & \vdots & \ddots & C_{N-1,N}\\
    0 &\ldots &C_{N, N-1} &\beta_{N}(z)\\
    \end{bmatrix},
\end{equation} 
having coupling strengths on the sub- and superdiagonal and the propagation constants on the diagonal. The solution to \eqref{eqn:coupled_waveguides} can be written in the form: 
    \begin{equation}
        \mathbf{A}(z)=V(z)\mathbf{A}(0),
    \end{equation}
where $V(z)=\exp\left(-i\int_0^{z}H(z')dz'\right)$ is the transfer matrix of the lattice. As the global phase accrued by the amplitudes in $\mathbf{A}(z)$ is usually irrelevant for measurable quantities, $N-1$ propagation constants $\beta_m(z)$ is enough to describe the phase shifts of $N$ field amplitudes in a layer. Therefore, without loss of generality, only $\beta$'s of first $N-1$ waveguides are considered variable, while the last one is fixed  through all layers: $\beta_N(z)=\beta_0$. In addition, it is convenient to eliminate $\beta_0$ from equations by the substitution: $\mathbf{A}(z)\rightarrow{}\exp(-i\beta_0z)\mathbf{A}(z)$. In this frame, the parameters relevant for field evolution are the propagating constant difference $\delta_m(z)=\beta_m(z)-\beta_0$ ($m=\overline{1,N-1}$). The same reasoning is applied to the phase shifts $\varphi^{(in)}_m$ and $\varphi^{(out)}_m$.

As in \eqref{eqn:coupled_waveguides}, the dependence of $H(z)$ on the propagation coordinate $z$ is due to the piecewise defined propagation constants, we accept the notation in which  $\delta_m(z)$ of the waveguide with index $m$ takes constant value $\delta_{mj}$ when $z$ falls into the layer with index $j$, i.e. $(j-1)l\le{}z<jl$ ($m=\overline{1,N-1}$, $j=\overline{1,N-1}$). Accordingly, the single system \eqref{eqn:coupled_waveguides} is split into $N-1$ systems, each described by a constant matrix $H_j$, and the transformation of the lattice is given by the product $V=V_{N-1}V_{N-2}\cdot\ldots\cdot{}V_{1}$, where
    \begin{equation}\label{eqn:lattice_layer}
        V_j=\exp(-iH_jl)
    \end{equation}
is the transformation of layer $j$ ($j=\overline{1,N-1}$).

The phase shifts added to the input and output of the scheme are  described by the diagonal matrices:
\begin{equation}\label{eqn:phase_shifts}
    P^{(s)} = \mathrm{diag}\left(e^{i\varphi^{(s)}_1},e^{i\varphi^{(s)}_2}, \dots ,e^{i\varphi^{(s)}_{N-1}},1\right).
\end{equation} 
where $s=in$ and $s=out$ for input and output, respectively.

%
%

\subsection{Relevant parameters of the architecture}

Using \eqref{eqn:lattice_layer} and \eqref{eqn:phase_shifts}, the transfer matrix of the $N$-channel scheme under study is given by the product of the transfer matrices, describing the layers of the lattice and the input and output phase shifts:
    \begin{equation}
        U=P^{(out)}V_{N-1}\cdot\ldots\cdot{}V_2V_1P^{(in)}.
    \end{equation}

To investigate the transformation capabilities of the proposed architecture, we consider the lattice with equal coupling coefficients ($C_{m,m+1}=C$), which is attained in practice by equal spacing between neighbouring waveguides. As a result, $C$ can be put out of matrices $H_j$ and  the dimensionless layer length  $\tilde{l}=Cl$ can be introduced in such a way that $H_jl=\tilde{H}_j\tilde{l}$, where matrix
    \begin{equation}
        \tilde{H}_j = \begin{bmatrix}
        \tilde{\delta}_{1j} & 1 & \ldots &   &0 \\
        1 &  \tilde{\delta}_{2j} &  &  &  \\
        \vdots &  & \ddots &  & \vdots\\
         &  &  & \tilde{\delta}_{N-1j} & 1\\
        0 &  & \ldots & 1 &  0\\
        \end{bmatrix}.
    \end{equation}
has the dimensionless entries $\tilde{\delta}_{mj}=\delta_{mj}/C$, i.e. the values of $\tilde{\delta}_{mj}$ are in the units of $C$.

\subsection{Optimization procedure}

To quantify the performance of the multichannel schemes, the following fidelity measure has been used
    \begin{equation}\label{eqn:fidelity}
        F(U,U_0)=\frac{|\text{Tr}(U^{\dagger}U_0)|^2}{\text{Tr}(U_0^{\dagger}U_0)\text{Tr}(U^{\dagger}U)}=\frac{1}{N^2}|\text{Tr}(U^{\dagger}U_0)|^2,
    \end{equation}
that compares the target transfer matrix $U_0$ and the actual transfer matrix $U$ realized by the lattice, where $N$ is the size of the lattice. Provided that the matrices $U$ and $U_0$ are equal up to a complex multiplier, the fidelity \eqref{eqn:fidelity} gets its maximum value of $F=1$.

The goal of the  analysis was to find a set of optimal lattice parameters (for each target transfer matrix $U_0$) that minimize the infidelity $1-F$. As no analytical solution is known to this task, we used numerical optimization procedure that looked for the global maximum of $F$ over the space of $N^2-1$ parameters $\left\{\delta_{mj}\right\}$, $\left\{\varphi^{(in)}_m\right\}$, $\left\{\varphi^{(out)}_m\right\}$  ($m=\overline{1,N-1};j=\overline{1,N-1}$). Namely, we used the basin-hopping global optimization algorithm implemented with the BFGS local optimizer provided by the SciPy python library. It should be noted that as an arbitrary length $\tilde{l}$ does not guarantee  optimal performance of the programmable scheme operating with the lowest possible infidelity, thus, we include the length in the parameter space of optimization as well.


\section{Results}

For the target unitary matrices $U_0$ we used random unitary matrices distributed with Haar measure \cite{Mezzadri2006} and the three specific types of matrices often finding applications in information processing: discrete Fourier transform (DFT) matrices \cite{Thyagarajan2019}, Hadamard matrices \cite{Horadam2010} and permutation matrices.

\begin{figure}[htbp]
\centering{\includegraphics[width=14cm]{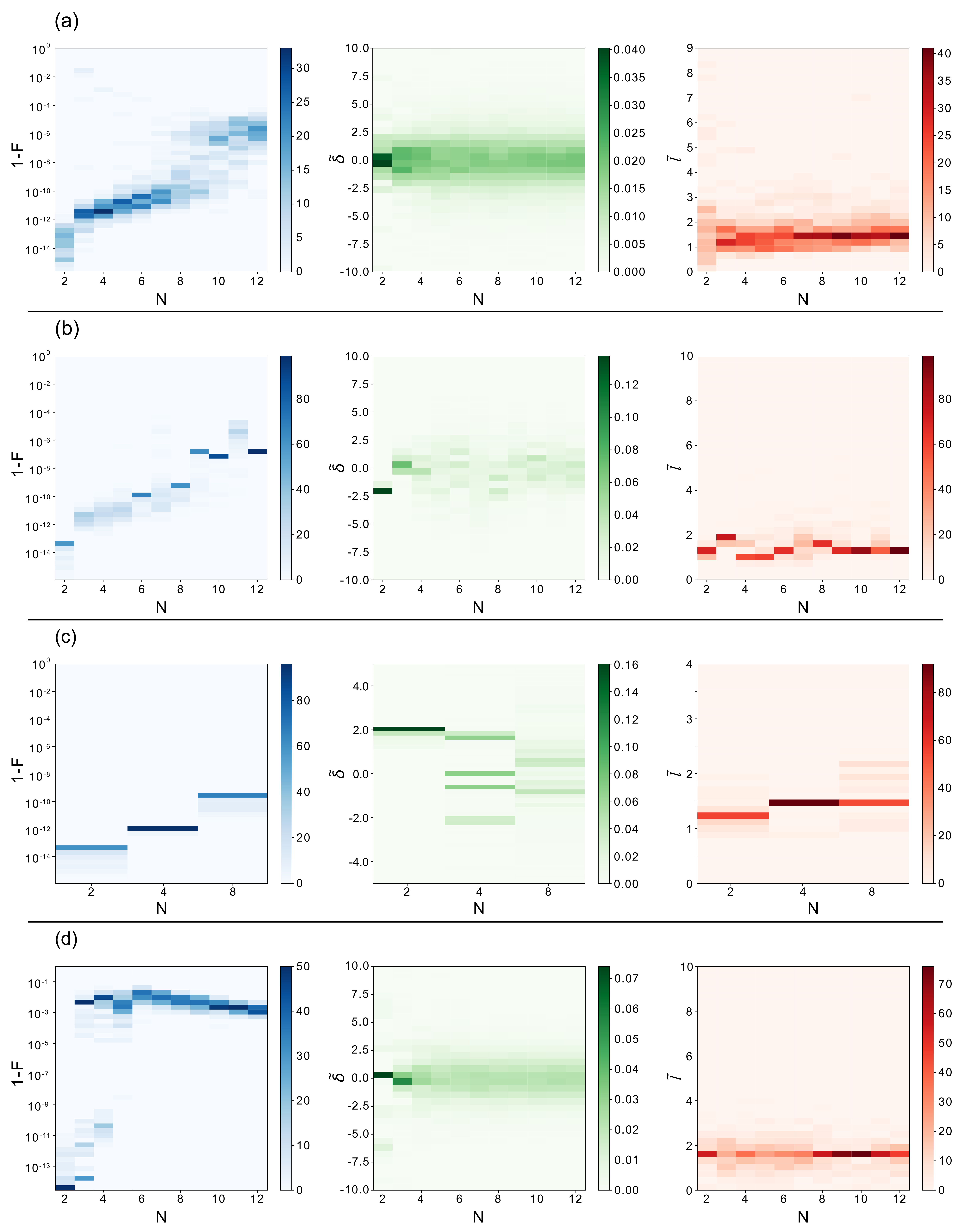}}
\caption{The histograms of infidelities $1-F$ (left), dimensionless  quantities $\tilde{\delta}_{mj}$ (middle) and layer length $\tilde{l}$ (right) in the cases of different target matrices: a) a set of $100$ Haar-random unitary matrices for each $N$, b) the DFT matrices at multiple runs of the optimization algorithm, c) Hadamard matrices at multiple runs of the optimization algorithm, d) a set of $100$ random permutation matrices.}\label{fig:histograms}
\end{figure}

Fig.~\ref{fig:histograms} summarizes the results of optimization. The histograms of the lowest achieved infidelities at different $N$ are shown on the left figures. In practice, $\delta_{mj}$ cannot take arbitrary large values due of the adopted assumptions and, more importantly, because they are limited by the strength of the effects exploited to tune the propagation constants, e.g. the thermo- or electro-optical ones (see Sec.~\ref{sec:discussion} for estimates). Therefore, we  monitor the values of $\delta_{mj}$ that implement target matrices, which distributions are shown in the middle part of Fig.~\ref{fig:histograms}. Also, the corresponding distributions of optimal layer lengths $\tilde{l}$ are shown on the right of Fig.~\ref{fig:histograms}.

The case of random matrices is shown in Fig.~\ref{fig:histograms}a. A set of $100$ Haar-randomly generated target matrices of $N$ ranging from $2$ to $12$ were generated. For larger $N$ the approximation precision drops which may be caused by the increased complexity of the optimized function landscape. Nevertheless the resulting configurations fit sample unitaries within the range of experimentally reachable precision.

For the DFT and Hadamard transformations for each particular matrix size we have run $100$ optimization procedures that enabled us explore the quality of optimization. The results are shown in Fig.~\ref{fig:histograms}b and Fig.~\ref{fig:histograms}c, respectively.
It should be noted that the optimizer finds different solutions with almost the same precision but with different lattice characteristics. The histograms at Fig.~\ref{fig:histograms}b and Fig.~\ref{fig:histograms}c show that the Fourier transformation accepts a much richer variety of equivalent solutions than the Hadamard transformation. 

Lastly, we performed optimization of the lattice to fit the permutation matrices (Fig.~\ref{fig:histograms}d). For this, a set of permutation matrices was generated prior to the optimization run. The lattice configuration was then tuned to reproduce the permutation matrices from this set. This class of transformations imposes the most strict requirements on the architecture parameters since the perfect result corresponds to the totally constructive interference at the given output channels.

\begin{figure}[h]
\centering{\includegraphics[width=14cm]{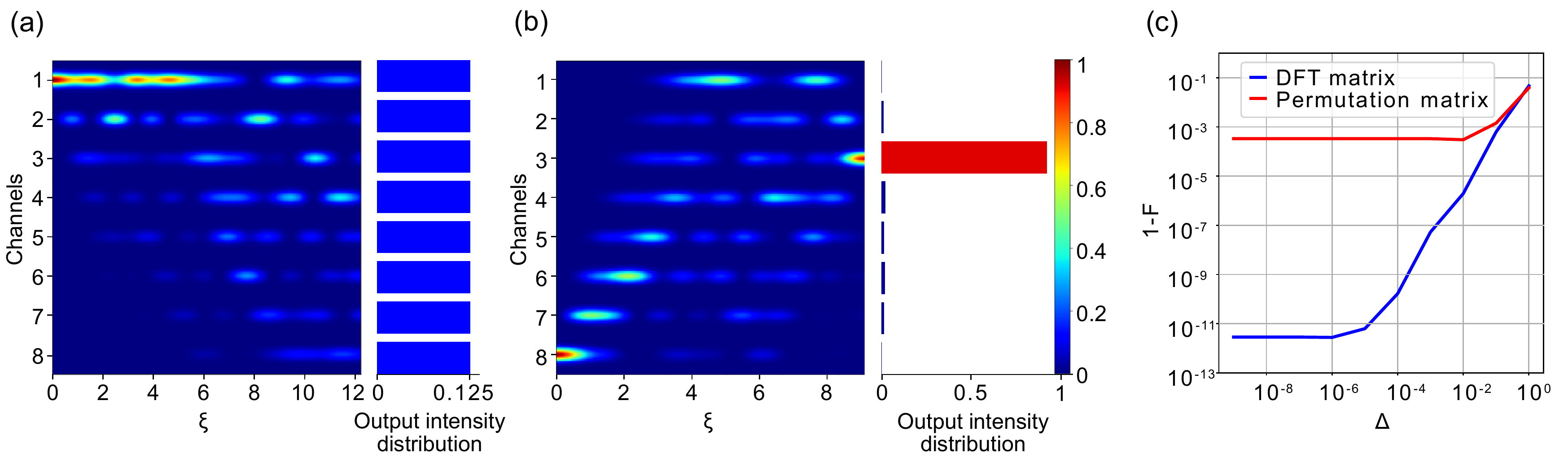}}
\caption{Simulated pattern of light propagating in two lattices  optimized for: a)  the DFT matrix  ($N = 8$, $1-F \sim 10^{-11}$) and b) for the permutation matrix ($N = 8$, $1-F \sim 10^{-3}$). In both cases single input channel is excited; $\xi=Cz$ is the normalized propagation coordinate. c) The illustration of the effect of inaccuracies in $\delta_{mj}$ and $l$ on these two  transformations quantified by the width of the error distributions  $\Delta=\Delta_{\tilde{\delta}}=\Delta_{\tilde{l}}$, where $\Delta_{\tilde{\delta}}$ and $\Delta_{\tilde{l}}$ are the distribution widths of  $\tilde{\delta}$ and $\tilde{l}$ inaccuracy, respectively.}\label{fig:inaccuracies}
\end{figure}

Fig.~\ref{fig:inaccuracies}a illustrates the field pattern that propagate in the lattice implementing  DFT of $8$ channels with optimized infidelity $1-F \sim 10^{-11}$ when excited in a single input channel. The corresponding field pattern for the permutation matrix of infidelity $1-F \sim 10^{-3}$ is in Fig.~\ref{fig:inaccuracies}b. Obvious solutions correspond to the much simpler light distributions for these kind of transformations. The optimizer finds the representative of the set of solutions which might correspond to a less trivial light propagation pattern.

To investigate the tolerance of the transformation quality against inaccuracy in $\delta_{mj}$ and $l$, we added normally distributed random errors into their the optimal values, characterized by the distribution widths $\Delta_{\tilde{\delta}}$ and $\Delta_{\tilde{l}}$. For simplicity, the widths were taken equal: $\Delta_{\tilde{\delta}}=\Delta_{\tilde{l}}=\Delta$. Then, the corresponding  fidelities have been calculated. Fig.~\ref{fig:inaccuracies}c compares the effect of errors in $\delta_{mj}$ and $l$ for the DFT and permutation matrices of  $N=8$. Here, the corresponding optimal infidelities in case of no errors are $\sim10^{-11}$ and $\sim10^{-3}$, respectively. Expectedly, elevating the degree of inaccuracy corrupts both transformations, however, while the error-free fidelity for the DFT matrix is much higher, it catches up with the one for the permutation matrix at $\Delta\sim10^{-1}$.



\section{Discussion}\label{sec:discussion}

Let us summarize the simulation results and figure out guidelines for the possible experimental implementation using currently available integrated photonic technologies. Here, we discuss schemes formed by weakly-guiding waveguide lattices, such as those fabricated by femtosecond direct laser writing and proton-exchanged waveguides. Therefore, in the estimates that follow we take the value of the coupling strength $C = 0.1$ mm$^{-1}$, which is typical for these lattices and small enough to ensure nearest-neighbour interactions. The results obtained above are saying that $-2.5\le\tilde{\delta}_{mj}\le2.5$ and $1.0\le\tilde{l}\le2.0$, so the layer length $l$ for these lattices is from $10$ to $20$ mm. The following is the estimates of propagation constant tuning related to the implementation of the programmable waveguide lattices by the thermo-optic and electro-optic effects:

\begin{enumerate}
    \item \textit{Thermo-optic tuning}. The thermo-optic phenomena is widely exploited in reconfigurable photonic chips, created by femtosecond direct laser writing, e.g. in fused silica. In the case of fused silica, usually chosen to be the material of the chip, the contrast of the refractive indices between the inscribed core ($n_{co}$) and the cladding ($n_{cl}$) amounts to $\sim 10^{-3}$. This requires the interwaveguide distance $d \sim 20$ $\mu m$ to achieve the coupling strengh  $C=0.1$  mm$^{-1}$ \cite{Szameit07}. In reprogramming by thermo-optics,  local heating of the material induce positive change of the refractive index, thus, we assume that $\delta$ covers the positive range from $0$ to $5C$ corresponding to values $0-0.5$ mm$^{-1}$ to meet the requirements. This yields the induced change of refractive index and the temperature raise to take the values $\Delta n = 6.4 \times 10^{-5}$ and $ \Delta T = 6.4$ $K$, respectively. The coupling constant modulation is of the order of $\Delta C/C$ = $\Delta n / (n_{co} - n_{cl})$ = $\lambda \Delta \beta / 2 \pi (n_{co} - n_{cl}) = 1.3\%$, which is small enough to not to take it into account, as was initially assumed. The femtosecond laser written thermo-optical tuners suffer from large amount of heat dissipation that come with the shift of refractive index to proper magnitudes and from intrinsically large cross-talks. Both these issues can be diminished by fabricating isolating trenches from the sides and the bottom of the waveguides and operating the chip in vacuum \cite{Ceccarelli2020}.
    
    \item \textit{Electro-optic tuning}. The modulation of propagation constants using electro-optic phemonena is utilized in  in proton-exchanged waveguides, created in lithium niobate crystals. For example, recent work by Youssry et al. \cite{Youssry20} explores the evanescently coupled waveguide lattice reconfigured by applying the electric field to individual waveguides of the lattice. Utilizing electro-optics rather than thermo-optics comes with the advantage of much lower cross-talks between the tuners and ultra-fast reprogrammability of the scheme. The coupling constant $C$ = $0.1$ mm$^{-1}$ is achieved at a distance of $d\approx 10$  $\mu m$ between adjacent waveguides. Notice that in reprogramming by electro-optics, both positive and negative changes of refractive indices are accessible by voltage polarity, hence, the tuning range $-2.5C\le\delta\le{}2.5C$ requires minimum and maximum values of $\delta=-0.25$ mm$^{-1}$ and $\delta=+0.25$ mm$^{-1}$. The magnitudes of the corresponding refractive index and voltage modulation is $\Delta n$ = 3.2 $\times$ 10$^{-5}$ and $\Delta{}V=6.4$ $V$. The calculated variations of coupling constants induced by maximal $\delta$ tuning are $\Delta{}C/C=0.5\%$, which is even smaller than that of the lattices reprogrammed by thermo-optics.

\end{enumerate}

Low-loss light coupling of the integrated photonic lattice circuit to the fiber arrays that have large spacing between the fibers can be a challenge, as it requires bent waveguides that occupy limited area on the chip. Currently, standard v-groove fiber optic assemblies are available with the minimal pitch of $127$ $\mu m$ between the fibers, thus, large fan-out section should be added to the lattice at the input and output sections. These rerouting sections might substantially increase the overall loss especially in the case of low-contrast waveguides. Meanwhile, non-standard pitch reducing solutions exist on the market with much lower fiber spacing of $30-40$ $\mu m$ and even down to $12$ $\mu m$ \cite{Kopp2012}, therefore, resolving the rerouting issue. It should be noted, that although we analyzed the lattice architecture by the model of coupled-mode theory with nearest-neighbour interactions, we expect the obtained results can be generalized for lithographic platforms within the framework of other more complex models.


\section{Conclusion}

In this work, we have considered the programmable architecture based on coupled waveguide lattices in order to investigate its capability to implement multiport transformations. Using the results of numerical simulation, we infer that the lattice architecture is suitable for efficient implementation of unitary transformations between $\sim{}10$ channels. The lattice architecture might substantially decreases the insertion losses in the devices operating by eliminating unnecessary sections used solely for routing the light on the chip. 

\section*{Acknowledgments}

This work was supported by Russian Foundation for Basic Research
grant No 19-52-80034. M.Yu.\,Saygin is grateful for support the Foundation for the Advancement of Theoretical Physics and Mathematics (BASIS).

\section*{Disclosures}

The authors declare no conflicts of interest.





\bibliography{sample}






\end{document}